\documentstyle[12pt]{ioplppt}

\newcommand{\be}{\begin{equation}}
\newcommand{\ee}{\end{equation}}
\newcommand{\bea}{\begin{eqnarray}}
\newcommand{\eea}{\end{eqnarray}}
\newcommand{\ket}{\rangle}

\newcommand{\bm}[1]{\mbox{\bf #1}}

\begin{document}
\jl{4}

\pagestyle{empty}
\begin{titlepage}

\title{Microscopic description of Gamow-Teller transitions
in middle {\em pf}--shell nuclei
by a realistic shell model calculation}[GT transitions
in middle {\em pf}--shell nuclei]

\author{H Nakada\dag\ and T Sebe\ddag}
\address{\dag\ Department of Physics, Chiba University,
 Inage-ku, Chiba 263, Japan}
\address{\ddag\ Department of Applied Physics, Hosei University,
 Koganei, Tokyo 184, Japan}

\begin{abstract}
GT transitions in $N=28\sim 30$ nuclei are studied
in terms of a large-scale realistic shell-model calculation,
by using Towner's microscopic parameters.
$B({\rm GT})$ values to low-lying final states are reproduced
with a reasonable accuracy.
Several gross properties with respect to the GT transitions
are investigated with this set of the wavefunctions
and the operator.
While the calculated total GT$^-$ strengths
show no apparent disagreement with the measured ones,
the calculated total GT$^+$ strengths are somewhat larger
than those obtained from charge-exchange experiments.
Concerning the Ikeda sum-rule,
the proportionality of $S_{\rm GT}$ to $(N-Z)$
persists to an excellent approximation,
with a quenching factor of 0.68.
For the relative GT$^-$ strengths
among possible isospin components,
the lowest isospin component gathers greater fraction
than expected by the squared CG coefficients
of the isospin coupling.
It turns out that these relative strengths
are insensitive to the size of model space.
Systematics of the summed $B({\rm GT})$ values
are discussed for each isospin component.
\end{abstract}

%\date{}

\maketitle

\end{titlepage}
\pagestyle{plain}
\setcounter{page}{1}

\section{Introduction}
\label{sec:intro}

Gamow-Teller (GT) transitions pose a challenging problem
for nuclear structure theories.
They play an important role in many astrophysical phenomena.
Middle {\em pf}--shell nuclei form a starting point
of the chain of the $s$-- and $r$--processes,
in which the GT transitions compete with the neutron capture.
Almost all heavier nuclei in nature
are synthesized via the $s$-- and/or $r$--processes.
Moreover, the GT transition rates
of middle {\em pf}--shell nuclei themselves
are significant inputs in the description
of supernova explosions.
On the other side, the GT transitions provide us
with a stringent test to nuclear many-body wavefunctions,
since they are sensitive to some details of the wavefunctions.

In the {\em sd}--shell region,
a realistic shell-model calculation is successful
in describing the GT transition strengths\cite{ref:BW85}.
Realistic shell-model approaches
to the GT transitions in the middle {\em pf}--shell nuclei
have been desired.
The extent of excitation out of the $0f_{7/2}$ orbit is crucial
to the GT transition rates in the middle {\em pf}--shell region.
However, in most of the shell-model wavefunctions
available so far,
the amount of the $0f_{7/2}$ excitation
has not been inspected sufficiently.
We have recently reported
one of the most successful realistic shell-model calculations
for the $N=28\sim 30$ nuclei\cite{ref:NSO94}.
Not only the energy levels,
but also the electromagnetic properties
are reproduced with the parameters derived
from microscopic standpoints.
We seem to have a reasonable amount of the leakage
out of $0f_{7/2}$ in these shell-model wavefunctions,
since the E2 and M1 properties are reproduced well
by the parameters on the microscopic ground.
In this article, we extend
this spectroscopically tested shell-model approach
to the GT transitions,
by using Towner's single-particle parameter-set\cite{ref:Towner}.

In addition to the $\beta$-decay strengths
between low-lying states,
the summed GT strengths have also been interested in.
Even with a reasonable quenching factor
for the GT transition operator,
there still remain discrepancies
between calculations and measurements
in some {\em pf}--shell nuclei.
While the final states of the GT$^+$ transitions
have a unique isospin value,
the GT$^-$ strength is distributed over a few isospin values,
because of the neutron excess.
In many cases, the $s$-- or $r$--process should be dominated
by the lowest isospin components.
Furthermore, the GT strengths with a specific isospin transfer
are required for some special topics.
For instance, in the double-$\beta$-decays
only the lowest isospin states can act as intermediate states.
It is also important to predict correctly the summed GT strength
with a specific isospin transfer.
However, the isospin distribution of the GT strength
has not been investigated so well.
Only recently it has become possible to acquire some information
from experiments\cite{ref:Fjt95},
on the isospin composition of the GT strength.
Whereas shell effects are expected
on the relative strength of each isospin component,
we have lacked systematic study based on realistic calculations.
We shall investigate the total GT strength
and its isospin partition,
as well as the decay strengths to low-lying states,
by using the realistic shell-model wavefunctions
in the middle {\em pf}--shell.

\section{Model space and GT operator}
\label{sec:apparatus}

The configuration space of the shell-model
calculation\cite{ref:NSO94} is as follows:
In the {\em pf}--shell on top of the $^{40}$Ca inert core,
we take the space consisting all of the $k\leq 2$ configurations,
where $k$ represents number of nucleons excited from $0f_{7/2}$.
Namely, $k$ is defined as
\be  (0f_{7/2})^{n_1-k}(0f_{5/2}1p_{3/2}1p_{1/2})^{n_2+k},
\label{eq:config} \ee
with $n_1=(Z-20)+8$ and $n_2=N-28$ 
for $20<Z\leq 28\leq N$ nuclei.
The $k\leq 2$ configuration space leads to the dimension
over 100\,000 in the $M$--scheme,
for several $N=30$ nuclei.
The Kuo-Brown hamiltonian\cite{ref:KBpf} is diagonalized
in this model space,
by using the code VECSSE\cite{VECSSE}.
In Reference \cite{ref:NSO94}, it has been shown
that the energy levels of the $N=28\sim 30$ nuclei
are reproduced within typical deviation of 0.3MeV.
The wavefunctions have been tested
via the E2 and M1 transition strengths and moments,
for which we have employed single-particle parameters
derived from microscopic standpoints.

The GT transition operator within the shell-model framework is
\be T({\rm GT}^\pm) = \sum_i  \left\{ g_{A}^{\rm eff}(nl)
 \sigma_i + g_{lA}^{\rm eff}(nl) l_i
+ g_{pA}^{\rm eff}(nl, n'l') [Y^{(2)}(\hat{\bm{r}}_i)
 \sigma_i]^{(1)} \right\} t_{\pm,i} , \label{eq:GTsp} \ee
where the sum runs over valence nucleons.
The $B({\rm GT})$ values are connected to the $ft$ values
through the following relation,
\be ft={{6170}\over{B({\rm GT})}} . \label{eq:ft} \ee
For free nucleons, we have $g_{A}=1.26$
and $g_{lA}=g_{pA}=0$.
It has been pointed out that nuclear medium effects
should be incorporated
into the $g$-parameters\cite{ref:Towner,ref:ASBH}.
We use Towner's microscopic single-particle 
parameter-set\cite{ref:Towner} as in the M1 case,
in which effects of the core-polarization (CP)
and the meson-exchange-currents (MEC)
on top of the $^{40}$Ca core
are taken into account.
No mass-number dependence is considered
for the $g$-parameters.
Apart from the finite values of $g_{lA}^{\rm eff}$
and $g_{pA}^{\rm eff}$,
the quenching for $g_{A}$ in the present parameter-set
is estimated to be $g_{A}^{\rm eff}/g_{A}=0.82$ ($0.81$)
for the $0f$ ($1p$) orbit.
In the operator of Equation (\ref{eq:GTsp}),
the MEC between a valence nucleon and the core
are taken into consideration.
Though the MEC between valence nucleons
lead to two-body operators,
their contribution is expected to be small,
as has been confirmed for the M1 quantities\cite{ref:NS94c}.
Remark that there is no adjustable parameters
in this calculation,
as in the calculation of the energy levels
and electromagnetic properties in Reference \cite{ref:NSO94}.

As has been stated already,
the $B({\rm GT})$ values are sensitive
to the $^{56}$Ni--core excitation
(i.e., the excitation out of the $0f_{7/2}$ orbit),
in the middle {\em pf}--shell region.
The GT transitions mainly concern
the nucleon-spin degrees-of-freedom.
The spin degrees-of-freedom will be active to a certain extent
for the middle {\em pf}--shell nuclei,
because the splitting of single-particle energies
makes the $0f_{7/2}$ orbit substantially occupied,
while its $LS$-partner $0f_{5/2}$ apt to be empty.
On the other hand, the interaction among nucleons favors
saturation of the nucleon-spin,
competing with the energy splitting.
Hence the calculated $B({\rm GT})$ values
are sensitive to the size of model space;
the calculated $B({\rm GT})$ values decrease
if larger amount of excitation out of $0f_{7/2}$ is involved.
The excitation from $0f_{7/2}$
to the upper {\em pf}--shell orbits
is important also for the E2 and M1 transitions.
The similarity of the M1-transition operator to the GT operator
has long been noticed\cite{ref:Towner,ref:ASBH}.
For the E2 transitions, the quadrupole collectivity,
which is a typical property
of the residual nucleon-nucleon interaction,
competes with the single-particle energy splitting.
Thereby the E2 transitions also have a significant correlation
with the excitation from $0f_{7/2}$ to the upper orbits.
As discussed in Reference \cite{ref:NSO94},
the E2 and M1 strengths are well described
by the present shell-model wavefunctions
with the parameters derived from microscopic standpoints.
Thus we have a reasonable amount of excitation out of $0f_{7/2}$
in the present wavefunctions,
and their application to the GT transitions may be promising.

It is commented that convergence for $k$
with the present shell-model hamiltonian is not evident.
On the other hand,
the spectroscopic test has a particular importance
in assessing the reliability of the wavefunctions.
Since it generally depends on the model space
whether the effective hamiltonian is appropriate or not,
such tests should be done for each set of space and hamiltonian.
Whereas the Kuo-Brown interaction has originally been developed
for the full {\em pf}--shell calculations,
the comprehensive reproduction
of the spectroscopic properties confirms
a certain reliability only of the present wavefunctions
obtained within the $k\leq 2$ space.
For this reason, we shall restrict ourselves in this paper
to the results extracted from the $k\leq 2$ wavefunctions,
putting aside the convergence problem.

\section{GT strengths to low-lying levels}
\label{sec:individual}

We hereafter restrict ourselves
to GT transitions from the ground states of the parent nuclei.
The GT-decay strength to individual low-lying state
is investigated first.
Both the initial and final states in this calculation
consist of the $k\leq 2$ configurations,
for which the wavefunctions have been well tested.
The calculated $B({\rm GT}^\pm)$ values are compared
with the measured ones\cite{ref:NDS},
in Tables~\ref{tab:GT-} and \ref{tab:GT+}.
This enables us to assess
how appropriate the present GT operator is.
Moreover, at the same time,
this comparison will be a further test
of the shell-model wavefunctions.
Several $B({\rm GT}^-)$ values are predicted
for possible $\beta$-decays,
in Table~\ref{tab:GT-}.
In Reference \cite{ref:NSO94},
calculated energy levels have been inverted in a few cases,
in making a correspondence to the observed ones,
based on the $B({\rm E2})$ and/or $B({\rm M1})$ values.
The inversion of the lowest two $({7\over 2})^-$ states
of $^{53}$Cr,
which is already taken into consideration in Table~\ref{tab:GT-},
is consistent with the GT strengths from $^{53}$V.
The $B({\rm GT}^+)$ values from $^{55}$Co
to $({9\over 2})^-$ states of $^{55}$Fe
suggest that the calculated $({9\over 2})^-_1$ level
should correspond to the observed $({9\over 2})^-_2$ state,
and {\it vice versa}.
Although this inversion has not been discussed
in Reference \cite{ref:NSO94}
and not considered in Table~\ref{tab:GT+},
it does not give rise to contradictions
to the electromagnetic properties.
Taking this into consideration, we find good agreement
between the calculated and measured GT strengths,
as far as the $B({\rm GT})$ values exceeding 0.01 are concerned.
As a general tendency, the present calculation slightly
underestimates the low-lying GT strengths.
However, except for a few transitions
from $^{56}$Ni and $^{57}$Ni,
those relatively large GT strengths are reproduced
within 70\% accuracy.
Furthermore, we have agreement in the order-of-magnitude
even for most other small GT strengths.
\begin{table}
\caption{\label{tab:GT-}
$B({\rm GT}^-)$ values.
The `Cal.' values are obtained
by the present shell-model calculation,
and the experimental data (Exp.) are taken
from Reference \protect\cite{ref:NDS}.}
\begin{indented}
\item[]\begin{tabular}{cc@{\qquad}cc@{\qquad\qquad}
r@{.}l@{\qquad}r@{.}l@{$\pm$}r@{.}l}
\br
 \multicolumn{2}{c@{\qquad}}{parent}
 & \multicolumn{2}{c@{\qquad\qquad}}{daughter}
 & \multicolumn{2}{c@{\qquad}}{Cal.}
 & \multicolumn{4}{c}{Exp.} \\
\mr
 $^{50}$Sc & $5_1^+$ & $^{50}$Ti & $4_1^+$ & 0&0001
  & \multicolumn{4}{c}{---} \\
 &&& $6_1^+$ & 0&022 & \multicolumn{4}{c}{---} \\
 &&& $4_2^+$ & 0&0020 & \multicolumn{4}{c}{---} \\
 $^{51}$Ti & $({3\over 2})_1^-$ & $^{51}$V & $({5\over 2})_1^-$
  & 0&081 & \multicolumn{4}{c}{---} \\
 &&& $({3\over 2})_1^-$ & 0&037 & \multicolumn{4}{c}{---} \\
 &&& $({3\over 2})_2^-$ & 0&0061 & \multicolumn{4}{c}{---} \\
 $^{52}$Ti & $0_1^+$ & $^{52}$V & $1_1^+$ & 0&361 & 0&56&0&04 \\
 &&& $1_2^+$ & 0&056 & \multicolumn{4}{c}{---} \\
 &&& $1_3^+$ & 0&106 & \multicolumn{4}{c}{---} \\
 $^{52}$V & $3_1^+$ & $^{52}$Cr & $2_1^+$
  &~~0&050 & 0&0616&0&0003 \\
 &&& $4_1^+$ & 0&0050 & 0&00022&0&00005 \\
 &&& $4_2^+$ & 0&0032 & 0&0075&0&0002 \\
 &&& $2_2^+$ & 0&0004 & 0&00306&0&00006 \\
 &&& $2_3^+$ & 0&0008 & 0&00048&0&00007 \\
 &&& $4_3^+$ & 0&011 & 0&007&0&003 \\
 &&& $3_1^+$ & 0&0033 & 0&0007&0&0004 \\
 $^{53}$V & $({7\over 2})_1^-$ & $^{53}$Cr & $({5\over 2})_1^-$
  & 0&156 & \multicolumn{4}{l}{0.15} \\
 &&& $({7\over 2})_1^-$ & 0&025 & \multicolumn{4}{l}{0.031} \\
 &&& $({7\over 2})_2^-$ & 0&0052
  & \multicolumn{4}{l}{$\sim$ 0.002} \\
 &&& $({9\over 2})_1^-$ & 0&0001 & \multicolumn{4}{c}{---} \\
 &&& $({9\over 2})_2^-$ & 0&0010 & \multicolumn{4}{c}{---} \\
\br
\end{tabular}
\end{indented}
\end{table}
\begin{table}
\caption{\label{tab:GT+}
$B({\rm GT}^+)$ values. The experimental data are taken
from Reference \protect\cite{ref:NDS}.}
\begin{indented}
\item[]\begin{tabular}{cc@{\qquad}cc@{\qquad\qquad}
r@{.}l@{\qquad}r@{.}l@{$\pm$}r@{.}l}
\br
 \multicolumn{2}{c@{\qquad}}{parent}
 & \multicolumn{2}{c@{\qquad\qquad}}{daughter}
 & \multicolumn{2}{c@{\qquad}}{Cal.}
 & \multicolumn{4}{c}{Exp.} \\
\mr
 $^{54}$Mn & $3_1^+$ & $^{54}$Cr & $2_1^+$ & 0&0036
  & \multicolumn{4}{l}{0.0039} \\
 $^{55}$Fe & $({3\over 2})_1^-$ & $^{55}$Mn & $({5\over 2})_1^-$
  & 0&0031 & 0&0062&0&0003 \\
 $^{55}$Co & $({7\over 2})_1^-$ & $^{55}$Fe & $({5\over 2})_1^-$
  &~~0&0068 & 0&0035&0&0003 \\
 &&& $({7\over 2})_1^-$ & 0&0046 & 0&00116&0&00006 \\
 &&& $({7\over 2})_2^-$ & 0&0027 & 0&0101&0&0006 \\
 &&& $({5\over 2})_2^-$ & 0&0028 & 0&0013&0&0002 \\
 &&& $({9\over 2})_1^-$ & 0&0079 & 0&0048&0&0004 \\
 &&& $({9\over 2})_2^-$ & 0&0021 & 0&010&0&001 \\
 $^{56}$Co & $4_1^+$ & $^{56}$Fe & $4_1^+$ & 0&00009
  & \multicolumn{4}{l}{0.00001} \\
 &&& $4_2^+$ & 0&000009 & 0&00014&0&00001 \\
 &&& $3_1^+$ & 0&0002 & 0&00066&0&00002 \\
 $^{56}$Ni & $0_1^+$ & $^{56}$Co & $1_1^+$ & 0&058
  & \multicolumn{4}{l}{0.25} \\
 $^{57}$Ni & $({3\over 2})_1^-$ & $^{57}$Co & $({3\over 2})_1^-$
  & 0&0079 & 0&0141&0&0006 \\
 &&& $({1\over 2})_1^-$ & 0&0039 & 0&0055&0&0001 \\
 &&& $({3\over 2})_2^-$ & 0&0027 & 0&0037&0&0002 \\
 &&& $({5\over 2})_1^-$ & 0&00002 & 0&0112&0&0005 \\
 &&& $({5\over 2})_2^-$ & 0&0028 & 0&00005&0&00001 \\
\br
\end{tabular}
\end{indented}
\end{table}

\section{Total GT strengths}
\label{sec:total}

The total GT strength is also a significant physical quantity
and has long been discussed.
There have been several shell-model calculations
for the middle {\em pf}-shell nuclei\cite{ref:SM-pf,ref:Lan95}.
We next apply the present shell-model approach to this issue.
The total GT strengths obtained
in the present shell-model calculation
are shown in Table~\ref{tab:GTtot}.
As is well-known, the total GT$^\pm$ strength
$\sum B({\rm GT}^\pm)$
is equal to the expectation value
of $T({\rm GT}^\mp)\cdot T({\rm GT}^\pm)$ in the initial state.
They are therefore a sort of ground-state property.
Although the final-state wavefunctions are not needed
to compute the total GT strengths,
we here discuss which configurations may come out
in the final states.
Since the initial state is comprised
of the $k\leq 2$ configurations
and the GT operator can excite a nucleon
from $0f_{7/2}$ to $0f_{5/2}$,
the GT$^-$ strength distributes
over the $k\leq 3$ configurations in the daughter nucleus,
for the transitions from the $N=28$ isotones.
The same holds for the Ni isotopes.
For the $N=29$ and $30$ isotones except $^{57}$Ni and $^{58}$Ni,
the $k=4$ configuration also emerges
in the final states of the GT$^-$ transitions.
This is because the lowest configuration is different
between the parent and daughter nucleus.
For instance, in $^{52}$Ti the $k=2$ configuration
implies $(0f_{7/2})^8 (0f_{5/2}1p_{3/2}1p_{1/2})^4$,
since $n_1=10$ and $n_2=2$ (see Equation (\ref{eq:config})).
The highest configuration generated by the GT$^-$ transition
is $(0f_{7/2})^7 (0f_{5/2}1p_{3/2}1p_{1/2})^5$.
This is the $k=4$ configuration of $^{52}$V,
the daughter nucleus,
because the lowest configuration shifts from $^{52}$Ti,
giving $n_1=11$, $n_2=1$.
Notice that this shift of $n_1$ and $n_2$
does not happen in the $N=28$ cases.
Thus the total GT$^-$ strengths of the $N=29$ and $30$ nuclei
are contributed by up to the $k=4$ configuration.
Conversely, the final-state configuration
in the GT$^+$ transition
is constrained into the $k\leq 2$ space,
for any of the nuclei under discussion.
The calculated eigenenergies of the ground state
may be disturbed,
if an admixture of the $k=3$ and $4$ configurations
is included explicitly.
However, the total GT$^\pm$ strengths are principally ruled
by the proton and neutron occupation numbers
of the $0f_{7/2}$ orbit\cite{ref:Aue87}.
Therefore, besides the convergence problem for $k$,
reliable total GT$^-$ strengths can be calculated
with the $k\leq 2$ ground-state wavefunctions,
as far as the wavefunctions contain
the excitation out of $0f_{7/2}$ properly.
\begin{table}
\caption{\label{tab:GTtot}
Total GT$^\pm$ strengths. The experimental data are taken
from References \protect\cite{ref:V51,ref:CX}.}
\begin{indented}
\item[]\begin{tabular}{cc@{\qquad\qquad}
r@{.}l@{\quad}r@{.}l@{\qquad}
r@{.}l@{$\pm$}r@{.}l@{\quad}r@{.}l@{$\pm$}r@{.}l}
\br
 \multicolumn{2}{c@{\qquad\qquad}}{parent}
 & \multicolumn{4}{c@{\qquad}}{Cal.}
 & \multicolumn{8}{c}{Exp.} \\
 && \multicolumn{2}{c}{$\sum B({\rm GT}^-)$}
 & \multicolumn{2}{c@{\qquad}}{$\sum B({\rm GT}^+)$}
 & \multicolumn{4}{c}{$\sum B({\rm GT}^-)$}
 & \multicolumn{4}{c}{$\sum B({\rm GT}^+)$} \\
\mr
 $^{49}$Sc & $({7\over 2})_1^-$ & 24&0 & 0&9
  &\multicolumn{4}{c}{---}&\multicolumn{4}{c}{---}\\
 $^{50}$Ti & $0_1^+$ & 21&2 & 1&5
  &\multicolumn{4}{c}{---}&\multicolumn{4}{c}{---}\\
 $^{51}$V & $({7\over 2})_1^-$ & 19&5 & 3&0
  & 20&0 & 4&0 &\multicolumn{4}{c}{---}\\
 $^{52}$Cr & $0_1^+$ & 17&4 & 4&3
  &\multicolumn{4}{c}{---}&\multicolumn{4}{c}{---}\\
 $^{53}$Mn & $({7\over 2})_1^-$ & 15&9 & 6&1
  &\multicolumn{4}{c}{---}&\multicolumn{4}{c}{---}\\
 $^{54}$Fe & $0_1^+$ & 14&3 & 7&7
  &~~12&4 & 3&0 &~~~5&6 & 0&8 \\
 $^{55}$Co & $({7\over 2})_1^-$ & 12&9 & 9&6
  &\multicolumn{4}{c}{---}&\multicolumn{4}{c}{---}\\
 $^{56}$Ni & $0_1^+$ &~~~~~11&4 &~~~~~11&4
  &\multicolumn{4}{c}{---}&\multicolumn{4}{c}{---}\\
\hline
 $^{50}$Sc & $5_1^+$ & 27&0 & 0&9
  &\multicolumn{4}{c}{---}&\multicolumn{4}{c}{---}\\
 $^{51}$Ti & $({3\over 2})_1^-$ & 24&2 & 1&3
  &\multicolumn{4}{c}{---}&\multicolumn{4}{c}{---}\\
 $^{52}$V & $3_1^+$ & 22&3 & 2&7
  &\multicolumn{4}{c}{---}&\multicolumn{4}{c}{---}\\
 $^{53}$Cr & $({3\over 2})_1^-$ & 20&1 & 3&8
  &\multicolumn{4}{c}{---}&\multicolumn{4}{c}{---}\\
 $^{54}$Mn & $3_1^+$ & 18&6 & 5&6
  &\multicolumn{4}{c}{---}&\multicolumn{4}{c}{---}\\
 $^{55}$Fe & $({3\over 2})_1^-$ & 16&9 & 7&2
  &\multicolumn{4}{c}{---}&\multicolumn{4}{c}{---}\\
 $^{56}$Co & $4_1^+$ & 15&5 & 9&1
  &\multicolumn{4}{c}{---}&\multicolumn{4}{c}{---}\\
 $^{57}$Ni & $({3\over 2})_1^-$ & 14&0 & 10&9
  &\multicolumn{4}{c}{---}&\multicolumn{4}{c}{---}\\
\hline
 $^{51}$Sc & $({7\over 2})_1^-$ & 30&1 & 0&8
  &\multicolumn{4}{c}{---}&\multicolumn{4}{c}{---}\\
 $^{52}$Ti & $0_1^+$ & 27&1 & 1&1
  &\multicolumn{4}{c}{---}&\multicolumn{4}{c}{---}\\
 $^{53}$V & $({7\over 2})_1^-$ & 24&9 & 2&2
  &\multicolumn{4}{c}{---}&\multicolumn{4}{c}{---}\\
 $^{54}$Cr & $0_1^+$ & 22&4 & 2&9
  &\multicolumn{4}{c}{---}&\multicolumn{4}{c}{---}\\
 $^{55}$Mn & $({5\over 2})_1^-$ & 20&8 & 4&6
  &\multicolumn{4}{c}{---}& 2&7 & 0&3 \\
 $^{56}$Fe & $0_1^+$ & 19&1 & 6&2
  & 15&7 & 3&8 & 4&6 & 0&5 \\
 $^{57}$Co & $({7\over 2})_1^-$ & 17&5 & 7&9
  &\multicolumn{4}{c}{---}&\multicolumn{4}{c}{---}\\
 $^{58}$Ni & $0_1^+$ & 15&8 & 9&5
  & 11&7 & 2&9 & 6&0 & 0&6 \\
\br
\end{tabular}
\end{indented}
\end{table}

The Ikeda sum-rule $S_{\rm GT}\equiv
\sum B({\rm GT}^-)-\sum B({\rm GT}^+)=3 g_A^2 (N-Z)$
is derived from the bare GT operator\cite{ref:IFF62}.
When we discuss the GT strengths within the $0\hbar\omega$ space,
a quenching factor for $g_A$ is sometimes introduced,
keeping $g_{lA}=g_{pA}=0$,
from phenomenological viewpoints.
Even in that case, $S_{\rm GT}$ is proportional to $(N-Z)$.
The present GT operator, however,
has finite values of $g_{lA}^{\rm eff}$ and $g_{pA}^{\rm eff}$,
because of the CP and MEC effects evaluated
from microscopic standpoints.
Hence $S_{\rm GT}$ is not exactly proportional to $(N-Z)$.
Figure \ref{fig:SR} depicts the $S_{\rm GT}$ values
obtained in the present microscopic calculation.
It turns out that the proportionality of $S_{\rm GT}$ to $(N-Z)$
is still maintained to an excellent approximation.
Quenching of $S_{\rm GT}$ is indicated,
resulting in 68\% of the bare value as a general trend.
The value of $(g_{A}^{\rm eff}/g_{A})^2$ is almost enough
to account for this quenching.

There have been attempts to extract total GT strengths
from charge-exchange (CX) reaction data
like $(p,n)$ and $(n,p)$\cite{ref:Tad87}.
The calculated total GT strengths are compared
with the available experimental values
of this kind\cite{ref:V51,ref:CX} in Table~\ref{tab:GTtot}.
Note that $g_A$ is included in the GT operator (\ref{eq:GTsp}),
associated with Equation (\ref{eq:ft}).
Most of the calculated $\sum B({\rm GT}^-)$ values
are consistent with the data,
although they are around the upper bound of the errors
of the experimental data.
For $\sum B({\rm GT}^+)$, the present calculation gives
somewhat larger values than the experiments.
This seems contradictory to the tendency
of slight underestimate of the individual low-lying GT strengths.
Possible reasons might be:
\begin{enumerate}
\item The GT strength might not fully be detected
in the CX experiments.
The level density is quite high in the GT resonance region
of the middle {\em pf}--shell,
which may make it difficult to measure all the GT strength
in this region.
In addition, a sizable amount of strength
within the $0\hbar\omega$ space
could be fragmented beyond the energy under observation.
There might be an ambiguity in the background subtraction
from the CX data,
as is argued for $^{48}$Ca\cite{ref:CPZ95}.
\item The CP and MEC mechanisms could somewhat differ
between the weak processes (i.e., $\beta$-decay
and electron capture) and the CX reactions,
although it is not likely that this gives rise to
a big quantitative difference.
\item There might be a problem
with the present shell-model wavefunctions.
However, the underestimate in the individual $B({\rm GT})$
suggests that the present wavefunctions and operator
involve too much leakage out of the $0f_{7/2}$ orbit,
whereas the total $B({\rm GT})$ appears opposite.
It is not obvious whether there are any better sets
of wavefunctions to dissolve this conflict.
One might worry about the influence
of the $k=3$ and $4$ configurations
on the final-state wavefunctions,
which is taken into account for the total GT$^-$ strengths,
while neglected for the low-lying strengths.
Notice that, however, the major conflict is present
in the GT$^+$ strengths,
for which we always consider the $k\leq 2$ model space,
regardless of the total or the low-lying strengths.
It should be mentioned here
that the total $B({\rm GT}^+)$ values
measured in the CX experiments are almost reproduced
by a recent full {\em pf}--shell calculation
by the shell-model Monte-Carlo (SMMC) method\cite{ref:Lan95}
with the so-called KB3 interaction,
despite the difficulty in performing precise spectroscopic tests
by the SMMC.
Further investigation will be required.
\end{enumerate}

A systematics of the total GT$^+$ strengths
has been suggested recently\cite{ref:LK94}, based on the CX data.
It has been pointed out that the summed GT$^+$ strengths
are nearly proportional to $(Z-20)\cdot (40-N)$.
Apart from the difference in absolute value
between the present calculation and the CX data,
we here test whether or not this systematics
is adaptable to the present shell-model results.
As is shown in Figure \ref{fig:GTsyst0},
the present results also obey
the $(Z-20)\cdot (40-N)$ systematics
to a good approximation.
The main difference between the data and the present calculation
is in the coefficient of $(Z-20)\cdot (40-N)$.
This fact suggests nearly a constant percentage
with which the GT$^+$ strengths are overlooked in the experiments
or overestimated in the calculation.
The calibration may probably be done by a single factor,
independent of nuclide, in this region.

\section{Isospin partition of GT$^-$ strengths}
\label{sec:isospin}

While final states of the GT$^+$ transition are constrained
to the lowest isospin,
the GT$^-$ transition strength yields an isospin distribution,
owing to the neutron excess.
A certain interest is being attracted
in the isospin distribution of the GT$^-$ strength.
As mentioned before, the GT strength
with a specific isospin transfer
may be key to some topical problems.
Recently, high-resolution CX experiments
have started to give quantitative information
on the GT$^-$ strengths
of each isospin component\cite{ref:Fjt95}.
Although there have been
a few shell-model calculations\cite{ref:V51},
no systematic study has yet been performed on this issue.
We next turn to a comprehensive study
on the isospin distribution of the GT$^-$ strengths.

In discussing the isospin composition of the GT$^-$ strength,
a simple argument based on the isospin algebra
has often been employed.
If we ignore effects of the shell structure,
the isospin partition is ruled
by the square of the isospin Clebsch-Gordan (CG) coefficients
$(T_0~T_0~1~-1|T_f~T_0-1)^2$\cite{ref:Ost92},
\bea &&\sum B({\rm GT}^-;T_0-1):\sum B({\rm GT}^-;T_0)
:\sum B({\rm GT}^-;T_0+1) \nonumber\\
&&~~~~~= {{2T_0-1}\over{2T_0+1}}:{1\over{T_0+1}}
:{1\over{(T_0+1)(2T_0+1)}} , \label{eq:CG} \eea
where $T_0$ denotes the isospin of the initial state
and $\sum B({\rm GT}^-;T_f)$ represents summed GT strengths
with the final isospin $T_f(=T_0-1,T_0,T_0+1)$.
Although this argument yields qualitative explanation
on some points,
such as the dominance of the $T_f=T_0-1$ component
in heavy nuclei,
it seems too simple for quantitative description
in the middle {\em pf}--shell and lighter region.
We shall look into the relative strengths
of each isospin component,
which have not been investigated well by realistic calculations.

In order to compute the relative strengths,
we first generate the GT$^-$ state
exhausting the transition strength from the initial state;
$|{\rm GT}^-\ket \propto T({\rm GT}^-) |i\ket$,
where $|i\ket$ stands for the initial state.
In this process, the $k\leq 3$ space is required
for the GT$^-$ state generated
from the $N=28$ isotones and the Ni isotopes,
while the $k\leq 4$ space should be considered
for the other $N=29$ and $30$ isotones.
Invoking the isospin-projection technique,
we separate each isospin component from the GT$^-$ state.
The relative strengths are obtained by the probabilities
of the respective isospin component in $|{\rm GT}^-\ket$.

In general, the shell structure increases
the relative strength of the low isospin component,
as will be clarified by the following argument.
We shall take $^{54}$Fe as an example, at first.
The ground state of $^{54}$Fe ($J^P=0^+$, $T_0=1$) has
the $(\pi 0f_{7/2})^6 (\nu 0f_{7/2})^8$ configuration
to the first approximation.
The GT$^-$ transition produces
the $(\pi 0f_{7/2})^7 (\nu 0f_{7/2})^7$
and $(\pi 0f_{7/2})^6 (\pi 0f_{5/2})^1
(\nu 0f_{7/2})^7$ configurations.
It is impossible, however, to form a $T_f=T_0+1=2$ state
with the $(\pi 0f_{7/2})^7 (\nu 0f_{7/2})^7$ configuration.
Moreover, there is no $1^+$ state with $T_f=T_0=1$
in the $(\pi 0f_{7/2})^7 (\nu 0f_{7/2})^7$ configuration.
The $T_f\geq T_0$ strengths are reduced by these mechanisms,
indicating an enhancement of the relative strength
of the $(T_0-1)$ component.
A similar discussion has been given for $^{58}$Ni
in Reference \cite{ref:Fjt95}.
The suppression of the $(T_0+1)$ fraction is expected
also for other nuclei,
not restricted to the middle {\em pf}--shell region,
via the same mechanism;
$T_f=T_0+1$ is forbidden in the configuration
where the proton having been converted by the GT$^-$ transition
occupies the same orbit as the initial neutron.
For the $N=28$ isotones,
the $T_0$ fraction is suppressed to a certain extent,
but less severely than the $(T_0+1)$ one;
for $T_f=T_0$, one or two angular momenta are not allowed
in some configurations with low seniority.
For the $Z<28<N$ nuclei including
most of the $N=29$ and $30$ isotones,
an argument about $k$,
the nucleon number excited out of $0f_{7/2}$
(see Equation (\ref{eq:config})),
clearly accounts for the shell-structure effect.
The initial state, namely the ground state of the parent nucleus,
is dominated by the $k=0$ configuration.
The GT$^-$ transition from this lowest configuration
generates the $k=0$, $1$ and $2$ configurations
of the daughter nucleus.
Note that the daughter nucleus has $Z\leq 28\leq N$.
With the $k=0$ configuration,
where the neutron $0f_{7/2}$ orbit is fully occupied,
only the lowest isospin can be formed.
Thus $T_f$ is restricted only to $(T_0-1)$
in the $k=0$ configuration.
In the $k=1$ configuration, the maximum possible isospin
is $T_f=T_0$.
The lack of the $k=0$ and $1$ configurations
hinders the $(T_0+1)$ component greatly,
and that of the $k=0$ configuration suppresses
the $T_0$ fraction moderately.
As a result, the $(T_0-1)$ fraction becomes larger
than estimated from Equation (\ref{eq:CG}).

This qualitative expectation is confirmed
in Figure \ref{fig:T-dist},
by the present realistic calculation.
In comparison with Equation (\ref{eq:CG}), 
the low isospin component has an enhancement
in its relative strength in any nucleus.
This trend is also consistent with the CX experiment
in $^{58}$Ni\cite{ref:Fjt95}.
In Figure \ref{fig:T-dist-sp}, the isospin partitions
in several different model spaces are compared.
Typical examples are taken
from an even-even ($^{56}$Fe), a proton-odd ($^{53}$V)
and a neutron-odd ($^{55}$Fe) nuclei.
The space $A$ is defined so that the initial-state wavefunction
should be comprised only of
the $(0f_{7/2})^{n_1} (1p_{3/2})^{n_2}$ configuration,
where $n_1$ and $n_2$ are defined
below Equation (\ref{eq:config}).
The space $B$ implies the $k=0$ space for the initial state,
whose wavefunction is obtained
by diagonalizing the Kuo-Brown hamiltonian.
The space $C$ indicates the present $k\leq 2$ model space,
for the initial-state wavefunction.
All possible configurations are considered for the final states.
The total GT$^-$ strengths appreciably depend
on the size of model space.
The difference between the spaces $A$ and $B$
is negligibly small,
as far as the same GT operator is employed.
The total $B({\rm GT}^-)$ value decreases considerably
if we use Towner's parameters,
into which the effects outside the $0\hbar\omega$ space
are incorporated.
The $k>0$ configurations reduce $\sum B({\rm GT}^-)$ further,
to about 60\% of the value
obtained in the space $A$ with the bare operator.
It turns out, on the other hand,
that the relative strengths are
quite insensitive to the truncation of the model space.
Although the isolation of the $0f_{7/2}$ orbit is relaxed
as enlarging the model space from $A$ to $C$,
the shell-structure effect on the relative strength
does not seem to become weaker.
The simplest wavefunctions in the space $A$
provides us with a seed of the shell effect,
and their isospin composition is preserved
in the higher configurations to be mixed, to a great extent.
This suggests that, if the total GT$^-$ strength is known,
we can get a sound and stable prediction
on the GT$^-$ strengths of definite isospin component
from wavefunctions with simple configurations.
It is also confirmed in the space $B$
that the modification of the GT operator from the bare one
hardly influences the relative strength.

Since we are dealing with as many as 24 nuclei
in the region $20<Z\leq 28\leq N\leq 30$,
it is possible to study systematics
of the summed GT$^-$ strengths for each isospin component,
based on the present shell-model calculation.
The principally active orbit in the ground state
is $0f_{7/2}$ for protons,
while $(0f_{5/2} 1p_{3/2} 1p_{1/2})$ for neutrons.
Therefore the systematics concerns multiple orbits,
not dominated only by $0f_{7/2}$.
The systematics of $S_{\rm GT}$ and $\sum B({\rm GT}^+)$
have been discussed already.
The $\sum B({\rm GT}^-;T_0+1)$ values are related
to the GT$^+$ strengths through the isospin algebra.
We here discuss the systematic behavior
of $\sum B({\rm GT}^-;T_0-1)$ and $\sum B({\rm GT}^-;T_0)$.
It is remembered that
the present calculation yields the same $(Z,N)$-dependence
for $\sum B({\rm GT}^+)$ as found in the CX data,
besides the overall coefficient.
Whether there is a missing of the summed GT strength
in the measurement
or an overcounting in the present calculation,
the study of systematics will be useful in calibrating
either of them,
as well as in speculating the summed GT strengths
of surrounding nuclei.
One might notice that not all of the four quantities,
$S_{\rm GT}$, $\sum B({\rm GT}^+)$,
$\sum B({\rm GT}^-;T_0-1)$ and $\sum B({\rm GT}^-;T_0)$,
are independent.
Only three of them can be independent,
because $S_{\rm GT}$ is equal
to $\sum B({\rm GT}^-) - \sum B({\rm GT}^+)$ by definition,
and $\sum B({\rm GT}^-;T_0-1)
= \sum B({\rm GT}^+)/[(T_0+1)(2T_0+1)]$.
It is, however, worth studying their systematics individually,
since the systematics is a matter of approximation.
Apart from $S_{\rm GT}$,
the systematics is investigated in an empirical manner
at present,
and its origin is not yet obvious.
Moreover, there is a notable difference in magnitude
among these quantities, as will be shown below.
Thus the precision of the systematics
may depend on the quantity under discussion.

The summed GT$^-$ strengths for the $T_f=T_0-1$
and $T_0$ components are presented in Figure \ref{fig:GTsyst1}.
The $(T_0-1)$ strengths are almost proportional to $2T_0=N-Z$,
with a trivial exception of the $T_0=1/2$ case
where $T_f=T_0-1$ is forbidden.
It is also found that
the $\sum B({\rm GT}^-;T_0-1)$ values are close to $S_{\rm GT}$.
The deviation of $\sum B({\rm GT}^-;T_0-1)$ from $S_{\rm GT}$
is less than 10\% in most cases.
This seems to be driven, to an appreciable extent,
by a cancellation between the GT$^-$ strengths
with $T_f\geq T_0$
and the GT$^+$ strength.
On the other hand, the summed $T_f=T_0$ strengths
behave in quite a different manner.
The $\sum B({\rm GT}^-;T_0)$ value decreases as $T_0$ increases,
and is roughly proportional to $1/(T_0+1)$,
which is just the squared isospin CG coefficient.
There should be some shell effects in the $T_f=T_0$ component,
as is stated above.
The systematics of $\sum B({\rm GT}^-;T_0)$ suggests
that the shell effects preserves the isospin dependence
of the squared CG coefficients approximately.
In this sense, the shell effects on $\sum B({\rm GT}^-;T_0)$
seem to contribute nearly uniformly in this region.
It is noted that
the $\sum B({\rm GT}^-;T_0+1)
=\sum B({\rm GT}^+)/[(T_0+1)(2T_0+1)]$
has another $(Z,N)$ dependence,
different from the strengths
of the $(T_0-1)$ and $T_0$ components.

According to the systematics,
the summed strengths $\sum B({\rm GT}^-;T_0-1)$
and $\sum B({\rm GT}^-;T_0)$ in this region
are functions only of $T_0$, to a relatively good approximation.
Since $\sum B({\rm GT}^-;T_0+1)$ amounts to less than 10\%
of the total GT$^-$ strength in the $T_0\geq 1$ cases,
the total strength is also a function of $T_0$
to a rough approximation,
although the function-form may be somewhat complicated.
As has been mentioned in connection with Equation (\ref{eq:CG}),
the $T_f=T_0-1$ component is dominant for large $T_0$.
It can be shown from the systematics
that the dominance of the $(T_0-1)$ component
grows even more rapidly than estimated
from Equation (\ref{eq:CG}), as $T_0$ increases.
The isospin algebra (Equation (\ref{eq:CG})) implies
that the ratio
$\sum B({\rm GT}^-;T_0-1)/\sum B({\rm GT}^-;T_0)$ rises
in $O(T_0)$ for increasing $T_0$.
The systematics implies, on the other hand,
that this ratio goes up in $O(T_0^2)$.
Note that the shell-structure effect is incorporated
into the simple functions representing the systematics,
in an effective manner.
In comparison with the estimate based on the isospin algebra,
the systematics implicates
that the shell-structure effect becomes the stronger
for the larger $T_0$,
and gives rise to the faster domination
of the low-isospin component.

Whereas the total GT strengths and their isospin distribution
have been studied systematically,
the energy distribution of the GT strengths
is left beyond the scope of this article,
except for the decay problems shown in Table~\ref{tab:GT-}
and \ref{tab:GT+}.
This is because,
despite the dispersion of the GT strength
over the $k=3$ and $4$ configurations,
spectroscopic test in the enlarged
(i.e., $k\leq 3$ or $k\leq 4$) space
has not been satisfactory yet.
A correction to the energy will be required,
when we take into account the admixture
of the $k=3$ and $4$ configurations.
We do not know a good way to do it at present,
and this point is left as a future problem.
It should be emphasized that, as stated already,
the total GT strength and their isospin partition
depend only on the ground-state wavefunction,
not concerning the energy correction.

\section{Summary}
\label{sec:summary}

GT transitions in $N=28\sim 30$ nuclei
have been investigated from a fully microscopic standpoint.
For individual low-lying GT strengths,
we have reproduced the relatively large $B({\rm GT})$ values
within 70\% accuracy
except for $^{56}$Ni and $^{57}$Ni.
While the calculated total GT$^-$ strengths
are in agreement with the measured ones
within the range of experimental errors,
the calculated total GT$^+$ strengths are somewhat larger
than those extracted from the CX experiments.
Although the GT strengths are calculated
with Towner's parameters incorporating medium effects,
the Ikeda sum-rule survives with a quenching factor of 0.68.
From the calculated GT$^-$ strengths
for respective $T_f$ component,
it is confirmed that low $T_f$ component gathers
larger relative strength
than expected by the squared isospin CG coefficients,
because of a shell effect.
Intriguingly, it turns out that this shell-structure effect
is hardly influenced by the size of model space;
the relative strength is insensitive to the extent
of the excitation out of $0f_{7/2}$.
The systematics of the summed $B({\rm GT})$ values
has also been discussed.
According to the study on the $\sum B({\rm GT}^+)$ systematics,
the discrepancy in the total GT strengths
between the present calculation and the CX experiments
seems to emerge with nearly a constant percentage.
The $T_0$ dependence of $B({\rm GT}^-;T_f)$ is quite different
among the $T_f=T_0-1$, $T_0$ and $T_0+1$ components.
Rapid growth of the $(T_0-1)$ component
relative to the $T_0$ one, for increasing $T_0$, is suggested.
Although the origin of the systematics is not yet clear enough,
this systematics might be useful to predict summed GT strengths
in surrounding nuclei without carrying out
elaborate computations.

\ack
The authors are grateful to Dr. Y. Fujita
for valuable discussions and comments.
They also thank Dr. W. Bentz for careful reading the manuscript.
A part of this work has been performed
within Project for Parallel Processing and Super-Computing
at Computer Centre, University of Tokyo.

\clearpage
\pagestyle{empty}
\section*{Figure Captions}
\begin{figure}
\caption{\label{fig:SR}
Calculated $S_{\rm GT}$ values as a function of $(N-Z)$.
Plus, circle and diamond symbols stand for
$N=28$, $29$ and $30$ nuclei, respectively.}
\end{figure}
\begin{figure}
\caption{\label{fig:GTsyst0}
Calculated total GT$^+$ strengths
as a function of $(Z-20)\cdot(40-N)$.
See Figure \protect\ref{fig:SR} for symbols.}
\end{figure}
\begin{figure}
\caption{\label{fig:T-dist}
Isospin composition of the GT$^-$ strengths:
$T_f=T_0-1$ (lightly shaded), $T_f=T_0$ (open)
and $T_f=T_0+1$ (darkly shaded) probabilities
in percentage.
Short sticks at the right of each column
indicate partitions due to the isospin algebra
(Equation (\protect\ref{eq:CG})).}
\end{figure}
\begin{figure}
\caption{\label{fig:T-dist-sp}
Model-space dependence of the isospin composition
of the GT$^-$ strength for $^{56}$Fe, $^{53}$V and $^{55}$Fe.
Each isospin component is expressed by the shaded area
as in Figure \protect\ref{fig:T-dist}.
The labels of the model spaces ($A$, $B$ and $C$)
are described in the text.
The GT operator is the bare operator
or the effective operator with Towner's parameter-set.
The cross symbols show the total $B({\rm GT}^-)$ values
relative to the value obtained in the space $A$
with the bare operator.}
\end{figure}
\begin{figure}
\caption{\label{fig:GTsyst1}
Systematic behavior of summed $B({\rm GT}^-)$ values
for $T_f=T_0-1$ and $T_0$ components.
Upper: $\sum B({\rm GT}^-;T_0-1)$ as a function of $2T_0=N-Z$.
Lower: $\sum B({\rm GT}^-;T_0)$ as a function
of $1/2(T_0+1)=1/(N-Z+2)$.
See Figure \protect\ref{fig:SR} for symbols.}
\end{figure}


\section*{References}
\begin{thebibliography}{99}
\bibitem{ref:BW85} Brown B A and Wildenthal B H 1985
 {\it At. Data Nucl. Data Tables} {\bf 33} 347 
\bibitem{ref:NSO94} Nakada H, Sebe T and Otsuka T 1994
 {\it Nucl. Phys.} {\bf A571} 467
\bibitem{ref:Towner} Towner I S 1987
 {\it Phys. Rep.} {\bf 155} 263
\bibitem{ref:Fjt95} Fujita Y {\it et al} 1996
 {\it Phys. Lett.} {\bf B365} 29
\bibitem{ref:KBpf} Kuo T T S and Brown G E 1968
 {\it Nucl. Phys.} {\bf A114} 241
\bibitem{VECSSE} Sebe T, Shikata Y, Otsuka T, Nakada H
 and Fukunishi N 1994
{\it VECSSE, Program library of Computer Centre,
 University of Tokyo}
\bibitem{ref:ASBH} Arima A, Shimizu K, Bentz W and Hyuga H 1988
 {\it Advances in Nuclear Physics vol.18}
 ed Negele J W and Vogt E (New York: Plenum) p 1
\bibitem{ref:NS94c} Nakada H and Sebe T 1994
 {\it Nucl. Phys.} {\bf A577} 203c
\bibitem{ref:NDS}  Peker L K 1984
 {\it Nucl. Data Sheets} {\bf 43} 481
\nonum Enchen Z {\it et al} 1985
 {\it Nucl. Data Sheets} {\bf 44} 463
\nonum Junde H {\it et al} 1989
 {\it Nucl. Data Sheets} {\bf 58} 677
\nonum Junde H 1991 {\it Nucl. Data Sheets} {\bf 64} 723
\nonum Bhat M R 1992 {\it Nucl. Data Sheets} {\bf 67} 195
\nonum Junde H 1992 {\it Nucl. Data Sheets} {\bf 67} 523
\nonum Junde H {\it et al} 1993
 {\it Nucl. Data Sheets} {\bf 68} 887
\bibitem{ref:SM-pf} Bloom S D and Fuller G M 1985
 {\it Nucl. Phys.} {\bf A440} 511
\nonum Muto K 1986 {\it Nucl. Phys.} {\bf A451} 481
\nonum Aufderheide M B, Bloom S D, Resler D A
 and Mathews G J 1993 {\it Phys. Rev.} {\bf C48} 1677
\nonum Auerbach N, Bertsch G F, Brown B A and Zhao L 1993
 {\it Nucl. Phys.} {\bf A556} 190
\nonum Caurier E, Mart\'{i}nez-Pinedo G, Poves A and Zuker A P,
 {\it Phys. Rev.} {\bf C52} R1736
\bibitem{ref:Lan95} Langanke K, Dean D J, Radha P B, Alhassid Y
and Koonin S E 1995 {\it Phys. Rev.} {\bf C52} 718
\bibitem{ref:Aue87} Auerbach N  1987
 {\it Phys. Rev.} {\bf C36} 2694
\bibitem{ref:IFF62} Ikeda K, Fujii S and Fujita J-I 1962
 {\it Phys. Lett.} {\bf 2} 169
\nonum Ikeda K, Fujii S and Fujita J-I 1963
 {\it Phys. Lett.} {\bf 3} 271
\bibitem{ref:Tad87} Taddeucchi T N {\it et al} 1987
 {\it Nucl. Phys.} {\bf A469} 125
\bibitem{ref:V51} Rapaport J {\it et al} 1984
 {\it Nucl. Phys.} {\bf A427} 332
\bibitem{ref:CX} Rapaport J {\it et al} 1983
 {\it Nucl. Phys.} {\bf A410} 371
\nonum El-Kateb S {\it et al} 1994
 {\it Phys. Rev.} {\bf C49} 3128
\nonum R\"{o}nnqvist T {\it et al} 1993
 {\it Nucl. Phys.} {\bf A563} 225
\bibitem{ref:CPZ95} Osterfeld F 1982
 {\it Phys. Rev.} {\bf C26} 762
\nonum Caurier E, Poves A and Zuker A P 1995
 {\it Phys. Rev. Lett.} {\bf 74} 1517
\bibitem{ref:LK94} Koonin S E and Langanke K 1994
 {\it Phys. Lett.} {\bf B326} 5
\bibitem{ref:Ost92} Osterfeld F 1992
 {\it Rev. Mod. Phys.} {\bf 64} 491
\end{thebibliography}
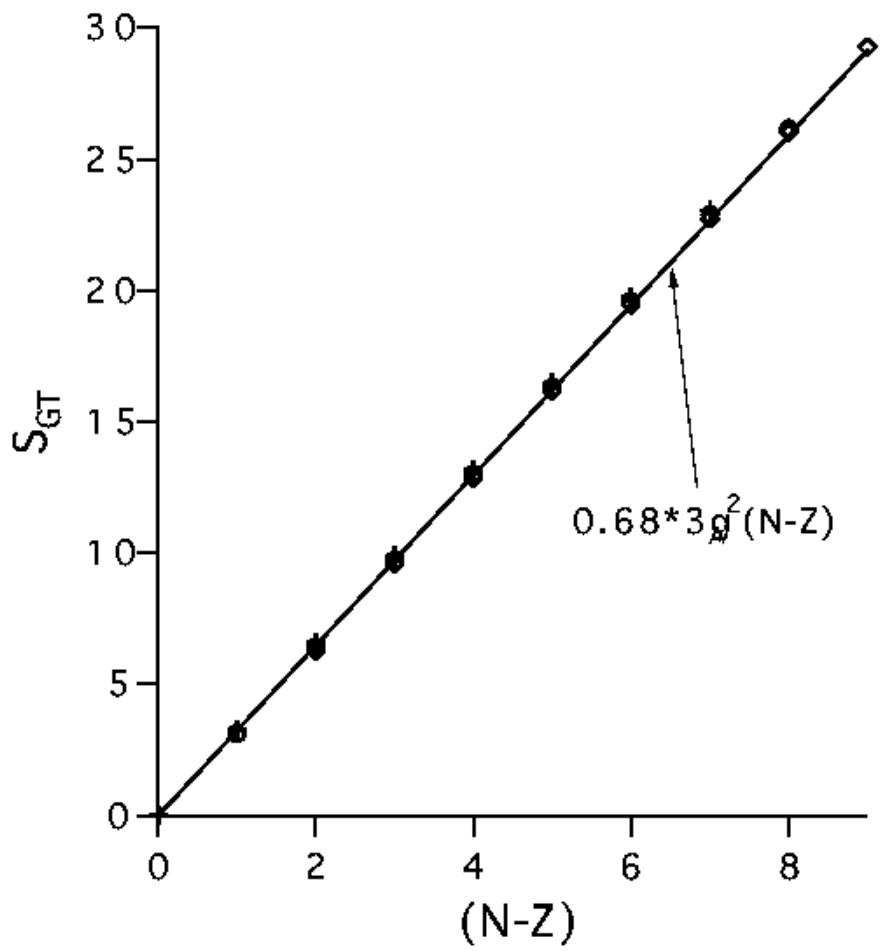
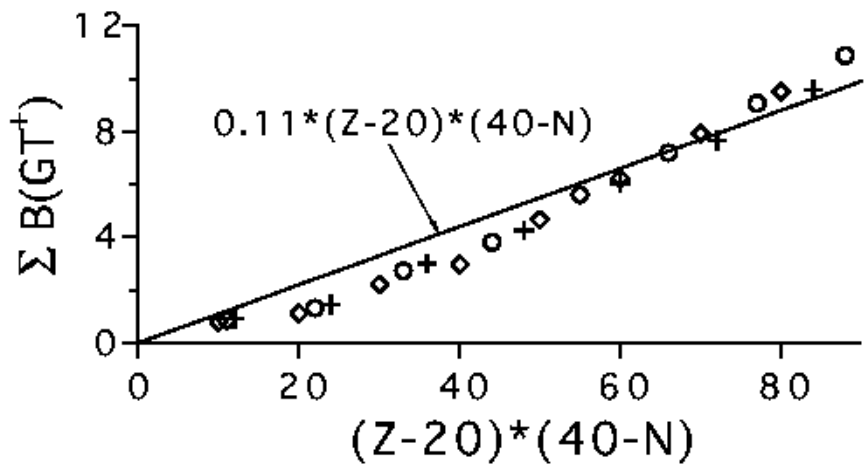
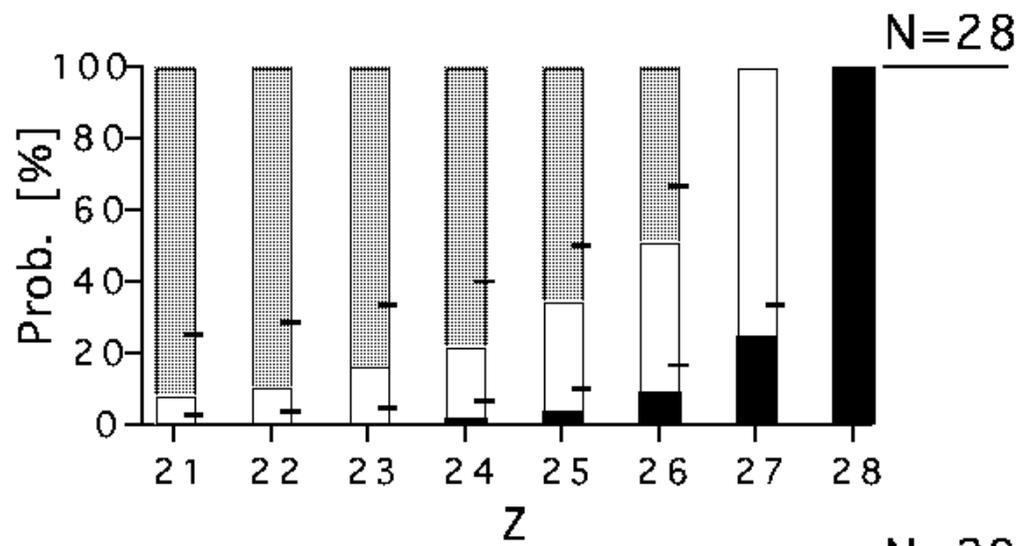
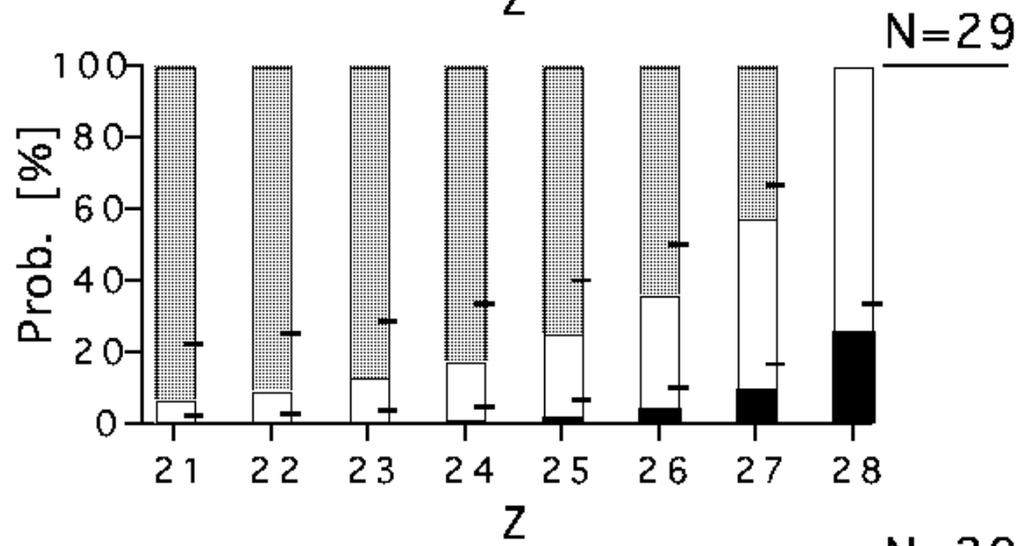
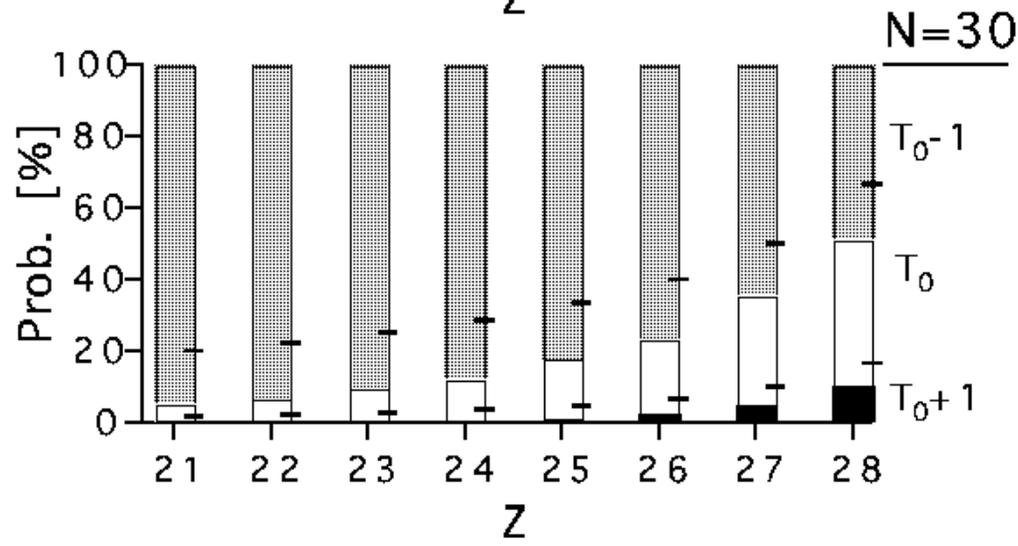
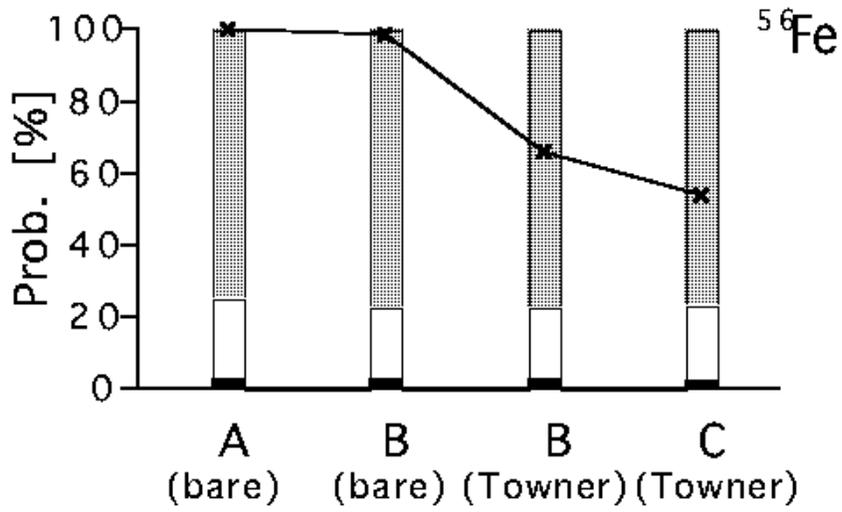
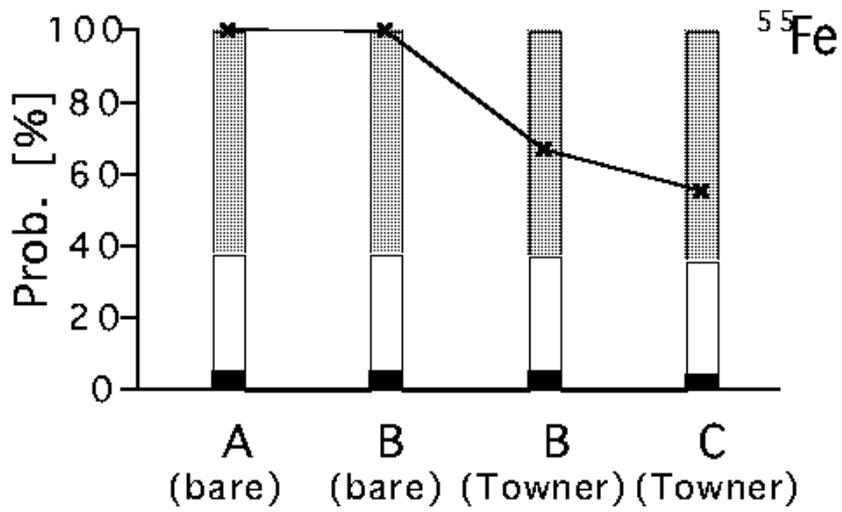
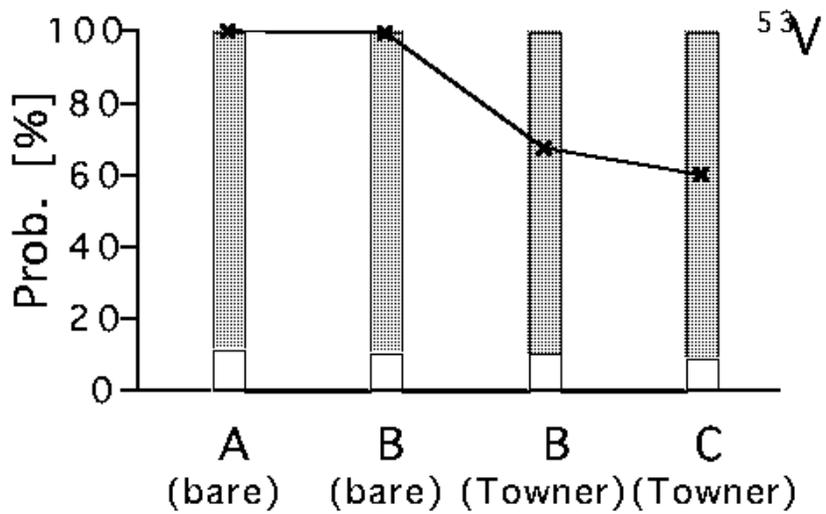
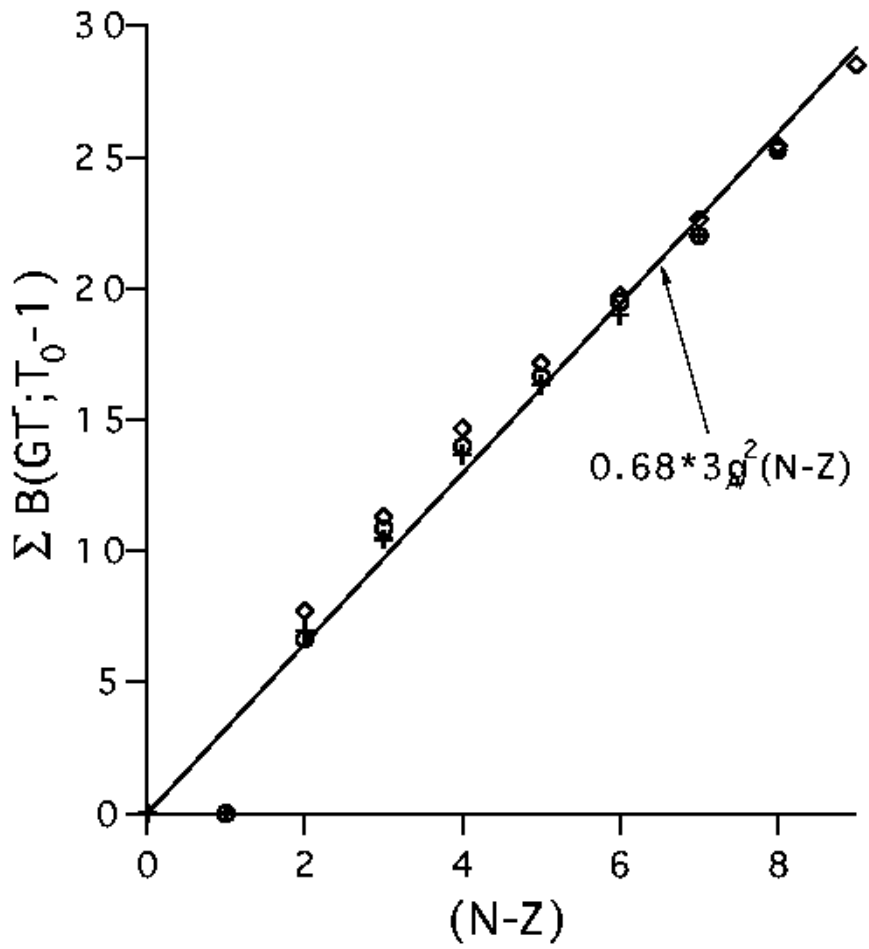
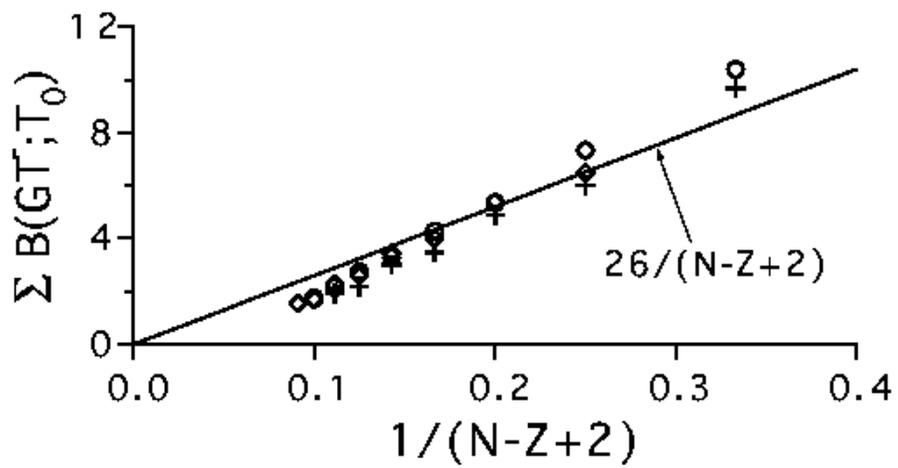
\end{document}